\documentclass[prd,twocolumn,showpacs,floatfix,amsmath,nofootinbib,amssymb,floatfix]{revtex4}
\usepackage{graphicx,color,dcolumn,booktabs,bm,multirow}
\usepackage{longtable,lscape}
\usepackage{txfonts}
\usepackage{overpic}
\usepackage{amssymb}
\usepackage{indentfirst}
\usepackage{feynmf}   %{feynmp}
\usepackage{slashed}  %for Feynman symbols
\usepackage{cases}
\usepackage{color}
\usepackage{multirow}
\usepackage{epstopdf}
\usepackage{graphicx,color,dcolumn,booktabs,bm}
\usepackage{epstopdf}
\usepackage[colorlinks,
            citecolor=blue,
            anchorcolor=red,
            menucolor=red,
            linkcolor=red,
            filecolor=red,
            runcolor=red,
            urlcolor=blue,
            frenchlinks=red]{hyperref}

\begin{document}

\title{A combined analysis of $Y(2240)$ in the $e^+ e^- \to K^+ K^-$ and $e^+ e^- \to K_S^0 K_L^0$ processes }
\author{Dian-Yong Chen$^{1,3}$}\footnote{Corresponding author}\email{chendy@seu.edu.cn}
\author{Jing Liu$^2$}
\author{Yu Bai$^1$}
\author{Chun-Hua Liu$^1$}
\author{Wen-Biao Yan$^4$}
\author{Zhao-Xia Meng$^5$}
\affiliation{
$^1$ School of Physics, Southeast University, Nanjing 210094, China\\
$^2$ School of Physics and electromechanical engineering, Hubei University of education, Wuhan 430205, China\\ 
$^3$ Lanzhou Center for Theoretical Physics, Lanzhou University, Lanzhou 730000, China\\
$^4$ School of Physical Science, University of Science and Technology of China, Hefei 230026, China\\
$^5$ School of Physics and Technology, University of Jinan, Jinan 250022, China
}

\date{\today}

\begin{abstract}
In the present work, we perform a combined analysis of the cross sections for $e^+ e^- \to K^+ K^-$ and $e^+ e^- \to K_S^0 K_L^0$ by including the interference between direct coupling process and resonance intermediate process. When setting the isospin factor as a free parameter, we can well reproduce the structures in the cross sections for $e^+ e^- \to K^+ K^-$ and $e^+ e^- \to K_S^0 K_L^0$, however, the isospin factor deviates with the isospin symmetry conservation predicted value of one. By fixing the isospin factor to be one, we find the cross sections for $e^+ e^-\to K^+ K^-$ can be accurately fitted, while the same model parameters fail to adequately reproduce the observed cross section for the process $e^+ e^- \to K_S^0 K_L^0$, which indicate that there should be contributions from other vector meson states, such as $\rho(2150)$ and $\omega(2220)$. More precise experimental measurement below 2.0 GeV and between $2.3\sim 2.5$ GeV may provide more information about the vector resonances in this energy range, which should be accessible for the BESIII collaboration.
 
\end{abstract}
\pacs{13.40.Gp, 13.66.Bc, 14.40.Cs}
\maketitle

%%%%%%%%%%%%%%%%%%%%%%%%%%%%%%%%%%%%%%%%%%%%%%%%%%%%%%
\section{Introduction}\label{sec1}
The strangeonium-like state $\phi(2170)$ was first reported by the BaBar Collaboration in $e^+ e^- \to \gamma_{\mathrm{ISR}}\phi f_0 (980)$ in 2006~\cite{BaBar:2006gsq}. The observed mass and width of $\phi(2170)$ are 
\begin{eqnarray}
	m_Y &=& (2175 \pm 10 \pm 15 )\ \mathrm{MeV}, \nonumber\\
	\Gamma_Y &=& (58 \pm 16 \pm 20)\ \mathrm{MeV},
\end{eqnarray}
respectively. In addition, the product of the branching ratio of $\phi(2170) \to \phi f_0(980)$ and the dilepton width was measured to be,
\begin{eqnarray}
	\Gamma_{ e^+ e^-}  \mathcal{B}_{Y\to \phi f_0}= (2.5 \pm 0.8 \pm 0.4)\ \mathrm{eV}.
\end{eqnarray} 
Furthermore, the measurements from BaBar Collaboration indicated that the events with no $f_0(980)$ candidate do not exhibit the structure of $\phi(2170)$ in $K^+K^- \pi^0 \pi^0$ mode, thus $\phi(2170)$ should decay to $\phi f_0(980)$ with a relatively large branching fraction. 

This strangeonium-like state had been confirmed by the BES Collaboration in the $J/\psi \to \eta \phi f_0(980)$ process, where the mass and width of $\phi(2170)$ were reported to be $(2186\pm 10 \pm 6)\ \mathrm{MeV}$ and $(65\pm 23\pm 17)\ \mathrm{MeV}$, respectively~\cite{BES:2007sqy}. Using the initial state radiation technique, the Belle Collaboration measured the cross sections for $e^+ e^- \to \phi \pi^+ \pi^-$ and $e^+ e^- \to \phi f_0(980)$, a structure corresponding to $\phi(2170)$ was reported, and the resonance parameters were measured to be $m_Y=(2079\pm 13^{+79}_{-28})\ \mathrm{MeV}$ and $\Gamma_Y=(192\pm 23^{+25}_{-61})\ \mathrm{MeV}$, respectively. The product of the branching fractions of $\phi(2170) \to \phi \pi^+ \pi^-$ and $\phi(2170) \to e^+ e^-$ was measured to be $\mathcal{B}_{Y\to \phi \pi^+ \pi^-} \mathcal{B}_{Y\to e^+ e^-}= (1.10 \pm 0.10 \pm 0.12)\times 10^{-7}$~\cite{Belle:2008kuo}. The large discrepancy of the resonance parameter from fits to the cross sections for  $e^+ e^- \to \phi \pi^+ \pi^-$ and $e^+ e^- \to \phi f_0(980)$ were supposed to be resulted from the assumptions on the background shape and the possible additional nearby resonances.

Since then, this states have further been observed by BaBar~\cite{BaBar:2011btv, BaBar:2007ceh, BaBar:2007ptr}, and BESIII~\cite{BESIII:2017qkh, BESIII:2020vtu, BESIII:2020gnc,BESIII:2020xmw, BESIII:2021bjn,BESIII:2022wxz,BESIII:2018ldc,BESIII:2021yam,  BESIII:2021aet, BESIII:2023xac} Collaborations in various channels. In Table~\ref{Tab:Exp}, We collect the experimental measurements of the resonance parameters of $\phi(2170)$ from different channels, where one can find that the measured mass and width are rather different in different channels. The latest PDG average of the resonance parameters of $\phi(2170)$ is~\cite{ParticleDataGroup:2024cfk}\footnote{It is worth mentioning that the resonance parameters of $\phi(2170)$ in PDG2018 were~\cite{ParticleDataGroup:2018ovx},
\begin{eqnarray}
m_Y &=& (2188\pm 10)\  \mathrm{MeV},\nonumber\\
\Gamma_Y &=& (83 \pm 12)\ \mathrm{MeV},\nonumber 
\end{eqnarray}
respectively, which is differ withe the latest PDG average by $2.1\sigma$ in mass and $1.1\sigma$ in width.
}, 
\begin{eqnarray}
m_Y &=& (2164 \pm 6) \ \mathrm{MeV},\nonumber\\
\Gamma_Y &=& (106^{+24}_{-18}) \ \mathrm{MeV},
\end{eqnarray}
respectively.

\renewcommand\arraystretch{1.4}
\begin{table*}[htb]
 \centering
 \caption{The resonance parameters of $\phi(2170)$ observed in various channels by BaBar, Belle and BES/BESIII Collaborations. \label{Tab:Exp}}
 \begin{tabular}{p{3.5cm}<\centering  p{3.5cm}<\centering p{5 cm}<\centering p{3cm}<\centering}
 %\hline\hline
 \toprule[1pt]
 \specialrule{0em}{1pt}{1pt}
   Mass (MeV) & Width (MeV) & Channel & Experiment\\
 %\hline\hline
 \midrule[1pt]
   $2175 \pm 10 \pm 15 $ & $58 \pm 16 \pm 20$&$e^+e^-\to \gamma_{\mathrm{ISR}} \phi f_0(980)$ & BABAR~\cite{BaBar:2006gsq}\\
   $2186\pm10\pm6$  & $65\pm23\pm17$  & $J/\psi \to \eta \phi f_{0}(980) $
    &BES~\cite{BES:2007sqy} \\
   $2125\pm22\pm10$  & $61\pm50\pm13$  & $e^+ e^- \to \gamma\phi\eta$
  &BABAR~\cite{BaBar:2007ceh}\\
   $2192\pm14$  & $71\pm21$  & $ e^+ e^-\to \gamma_{\mathrm{ISR}}  K^+ K^-\pi^+ \pi^-$
  &BABAR~\cite{BaBar:2007ptr}  \\
   $2169\pm20$  & $102\pm27$  & $ e^+ e^-\to \gamma_{\mathrm{ISR}}  K^+ K^-\pi^0 \pi^0$
  &BABAR~\cite{BaBar:2007ptr}\\
   $ 2079\pm13_{-28}^{+79}$  & $192\pm23_{-61}^{+25} $  & $e^+ e^- \to \gamma_{\mathrm{ISR}} \phi \pi^+ \pi^- $
  &Belle~\cite{Belle:2008kuo} \\
   $2200\pm6\pm5$  & $104\pm15\pm15$    & $ J/\psi   \to \eta \phi \pi^+ \pi^- $
  &BESIII~\cite{BESIII:2014ybv}\\
     $2135\pm8\pm9 $  & $ 104\pm24\pm12$  & $e^+ e^- \to \eta\phi f_{0}(980) $
  &BESIII~\cite{BESIII:2017qkh}\\
   $2126.5 \pm 16.8 \pm 12.4$  & $106.9 \pm 32.1 \pm 28.1$  & $ e^+ e^-   \to K^+ K^-\pi^0 \pi^0 $
  &BESIII~\cite{BESIII:2020vtu}\\
    $2177.5 \pm 4.8\pm 19.5$  & $149.0 \pm 15.6 \pm 8.9 $  & $ e^+ e^-   \to\phi \eta^{\prime} $
  &BESIII~\cite{BESIII:2020gnc}
  \\
      $2179\pm21\pm3$  & $89\pm  28\pm 5$  & $ e^+ e^-   \to\omega\eta $
  &BESIII~\cite{BESIII:2020xmw}
  \\
    $2163.5\pm6.2\pm3.0$  & $31.1_{-11.6}^{+21.1}\pm1.1$  & $e^+ e^- \to \phi \eta $
  &BESIII~\cite{BESIII:2021bjn} \\
 $2178\pm 20 \pm 5$ & $140\pm 36\pm 16$ & $e^+e^-\to \phi \pi^+ \pi^-$ & BESIII~\cite{BESIII:2021aet}\\	
  $2190\pm19\pm37$  & $191\pm28\pm60$  & $ e^+ e^-   \to K^+K^-\pi^0 $
  &BESIII~\cite{BESIII:2022wxz}\\
 $2164.7 \pm 9.1 \pm 3.1$ & $32.4 \pm 21.0 \pm 1.8$ & $e^+ e^- \to K_S K_L \pi^0 $ & BESIII~\cite{BESIII:2023xac}\\
\midrule[1pt]
 $ 2239.2 \pm 7.1 \pm 11.3$  & $ 139.8\pm 12.3 \pm 20.6$  & $ e^+ e^-   \to K^+ K^- $
  &BESIII~\cite{BESIII:2018ldc}\\
   $2273.7\pm5.7\pm19.3$  & $86 \pm 44 \pm 51$  & $ e^+ e^-   \to K^{0}_{S}K^{0}_{L} $
  &BESIII~\cite{BESIII:2021yam}
  \\
 \bottomrule[1pt]
 %\hline\hline
 \end{tabular}
  \centering
 \end{table*}

After the observation of $\phi(2170)$, it aroused theorists' great interests since $\phi(2170)$ was only observed in the hidden-strange channels at the early stage, such as $\phi \pi^+ \pi^-$, which is similar to the case of $Y(4260)$ in $\pi^+ \pi^- J/\psi$~\cite{BaBar:2005hhc}, $Y(4360)$ in $\pi^+ \pi^- \psi(2S)$~\cite{BaBar:2006ait}, and $Y_b(10890)$ in $\pi^+ \pi^- \Upsilon(nS),\ (n=1,2,3)$~\cite{Belle:2007xek}. Thus, $\phi(2170)$ was considered as the strange counterpart of $Y(4260)/Y(4360)$ and $Y_b(10890)$, and the exotic interpretations, such as tetraquark, hadronic molecular and hybrid interpretations, have been proposed. The estimations by using the QCD sum rule approach~\cite{Wang:2006ri,Jiang:2023atq,Agaev:2019coa}, the non-relativistic Hamiltonian frame~\cite{Drenska:2008gr}, and the flux-tube model~\cite{Deng:2010zzd} indicated that $\phi(2170)$ could be a compact tetraquark state. The evaluations in the frame of both flux-tube model and constituent gluon model  favor the hybrid interpretations~\cite{Ding:2006ya, Ding:2007pc}, while the non-perturbative Lattice and QCD Gaussian sum rule estimations disfavor the hybrid assignment for $\phi(2170)$~\cite{Dudek:2011bn, Ho:2019org}. In addition, it's worth mention ing that in the vicinity of $\phi(2170)$, there are abundant thresholds of a pair of hadrons, such as $\Lambda \bar{\Lambda}$, $\phi f_0(980)$, thus, some hadronic molecular interpretations have been proposed, such as the $\Lambda \bar{\Lambda}$ baryonium~\cite{Klempt:2007cp, Dong:2017rmg}, the $\phi K \bar{K}$ resonance~\cite{MartinezTorres:2008gy}. Besides the above exotic interpretations, $\phi(2170)$ had also been tried to be categorized as a conventional strangeonium, such as $3^3S_1$~\cite{Barnes:2002mu} or $2^3D_1$~\cite{Ding:2006ya,Wang:2012wa} states.

From the perspective of the decay behavior, the estimations in Ref. ~\cite{Jiang:2023atq} indicated that the $ss\bar{s}\bar{s}$ compact tetraquark states should dominant decay into two meson with $s\bar{s}$ quark components, such as $\phi \eta^{(\prime)}$, $\phi f_0(980)$ and $h_1(1415)\eta^{(\prime)}$, while the $\phi f_0$, $\phi \eta^{(\prime)}$ should be the dominant decay modes of $su\bar{s}\bar{u}$ tetraquark state~\cite{Agaev:2019coa}. In the strangeonium hybrid scenario, $\phi(2170)$ should dominantly decays into open-strange channels, such as $K\bar{K}$, $K^\ast \bar{K}$, $K^\ast \bar{K}^\ast$, $K_1(1270)\bar{K}$, and $K_1(1400)K$~\cite{Ding:2006ya}. The estimations in Ref.~\cite{Ding:2007pc} also indicated that  the decay patterns of strangeonium hybrid and $2^3D_1$ strangeonium are very different, the open strange decay modes were suggested to distinguish these two picture. Thus, searching the signal of $\phi(2170)$ in the open strange channels are crucial.

  In 2018, the BESIII Collaboration measured the cross sections for $e^+ e^- \to K^+ K^-$ at $\sqrt{s}=2.0 \sim 3.08$ GeV with the best precision achieved so far, no signal of $\phi(2170)$ was observed, but a new structure, named $X(2240)$, was reported with the resonance parameters to be~\cite{BESIII:2018ldc},
\begin{eqnarray}
m_{X(2240)} &=& 2239.2 \pm 7.1 \pm 11.3 \ \mathrm{MeV},\nonumber\\
\Gamma_{X(2240)} &=& 139.8 \pm 12.3 \pm 20.6 \ \mathrm{MeV},
\end{eqnarray}
which is differ from the PDG average mass of the $\phi(2170)$ state by more than $5 \sigma$ \footnote{The PDG average mass of $\phi(2170)$ was $(2188 \pm 10)$ MeV in 2019~\cite{ParticleDataGroup:2018ovx}. Then the measured mass of $X(2240)$ is only differ from the old PDG average mass of $\phi(2170)$ by about $3 \sigma$ as indicated in Ref.~\cite{BESIII:2018ldc}.}, and also differs from most individual experiments as listed in Table~\ref{Tab:Exp}.

In addition to the $e^+ e^- \to K^+ K^-$ process, the BESIII Collaboration also measured the cross sections for $e^+ e^- \to K_S^0 K_L^0$~\cite{BESIII:2021yam} at $\sqrt{s}=2.0 \sim 3.08$ GeV. A resonant structure around 2.2 GeV was reported, and the resonance parameters were, 
\begin{eqnarray}
	m &=&(2273.7\pm 5.7 \pm 19.3) \ \mathrm{MeV}, \nonumber\\
	\Gamma &=& (86\pm 44\pm 51) \ \mathrm{MeV},
\end{eqnarray}
respectively, which are consistent with the one observed in the cross sections for $e^+ e^- \to K^+ K^-$ within $2\sigma$ in mass and $1\sigma $ in width. In addition, the product of the dilepton decay width of $X(2240)$ and the branching fraction of $X(2240)\to K_S^0 K_L^0$ was measured to be,
\begin{eqnarray}
	\Gamma_{e^+ e^- } \mathcal{B}_{X\to K_S^0 K_L^0} =(0.9 \pm 0.6 \pm 0.7) \ \mathrm{eV}.
\end{eqnarray}
It should be noted that the cross sections for $e^+ e^- \to K^+ K^-$ are about one order of magnitude larger than those for $e^+ e^- \to K_S^0 K_L^0$. In the continuum region, this phenomena have been investigate in Ref.~\cite{Jin:2019gfa}. However, large discrepancy appear even in the resonance region, and one has $\Gamma_{X\to K^+ K^-}\sim 10 \times \Gamma_{X\to K_S^0 K_L^0}$, which  strongly violate the isospin symmetry. 

\begin{figure}[t]
\centering
\begin{tabular}{cc}
\scalebox{0.4}{\includegraphics{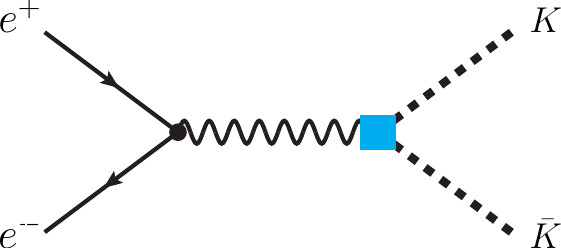}}& 
\scalebox{0.4}{\includegraphics{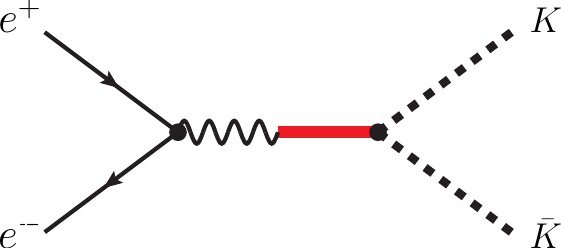}}\\
(a)& (b)
\end{tabular}	
\caption{(Color online.) The mechanisms working in $e^+ e^- \to K \bar{K}$ process. Diagram (a) is the direct coupling process, where the cyan diamond indicate the form factor of kaon in timelike region. Diagram (b) is the resonance intermediate process, where the red line indicate the intermediate vector meson. \label{Fig:Mech}}
\end{figure}

From the BESIII measurements of the cross sections for $e^+ e^- \to K^+ K^-$ and  $e^+ e^- \to K_S^0 K_L^0$, two interesting puzzles arise. The first one is about the relationship of $X(2240)$ and $\phi(2170)$, specifically, whether  $X(2240)$ and $\phi(2170)$ are the same state. The second puzzle is how to understand the large isospin violation in the processes of $X(2240)\to K^+ K^-$ and $X(2240) \to K_S^0 K_L^0$. As indicated in Ref.~\cite{Chen:2020xho} that the $e^+ e^- \to K^+ K^-$ process could occur via two different mechanism, which are the direct coupling process and the resonance intermediate process. As shown in Fig.~\ref{Fig:Mech}-(a), the $e^+e^-$ annihilate to virtual photon and the photon couples to the $K^+K^-$ directly in the former mechanism, in this process the time-like form factor of kaon is involved.  In the second mechanism, the virtual photon couples to a vector meson and the vector meson decays into $K^+ K^-$ as shown in Fig.~\ref{Fig:Mech}-(b). With the interferences between these two kinds of mechanism, the experimental data could be well described with resonance parameters of $X(2240)$ to be $m=(2197.4 \pm 4.4)\ \mathrm{MeV}$ and $\Gamma=(75.6\pm 7.2) \ \mathrm{MeV}$, which are consistent with most individual experiments, thus we conclude that $X(2240)$ and $\phi(2170)$ should be the same state. In the present work, we expand the interference picture in Ref.~\cite{Chen:2020xho} to the process $e^+ e^- \to K_S^0 K_L^0$, and by the  combined analysis of the cross sections for $e^+ e^- \to K^+ K^-$ and $e^+ e^- \to K_S^0 K_L^0$, we can also obtain the ratio of the widths of $K^+ K^-$ and $K_S^0 K_L^0$ channels, which may shield light on the isospin breaking puzzle of $X(2240)$.

This work is organized as follows, after introduction, we preset the interferences picture of $e^+ e^- \to K\bar{K}$ in the following section. In Section \ref{Sec:Num}, we present the fit results and the related discussions, and the last section is devoted to a short summary.

\section{Interference mechanism of $e^+ e^- \to K\bar{K}$ \label{Sec:Mech}}
As indicated in Ref.~\cite{Chen:2020xho}, there are two kinds of mechanisms working in the $e^-(p_1) e^+(p_2) \to K(p_3) \bar{K}(p_4)$, and the Feynman diagrams are presented in Fig.~\ref{Fig:Mech}. Here, we construct the amplitudes corresponding to the diagrams in Fig.~\ref{Fig:Mech} by the effective Lagrangian approach. The effective Lagrangian of $KK\gamma$ is,
\begin{eqnarray}
\mathcal{L}_{KK\gamma} &=& ie A^\mu (\bar{K} \partial_\mu K-\partial_\mu \bar{K} K), \label{Eq:Lag1}
\end{eqnarray}
then the amplitude for the direct coupling process corresponding to Fig.~\ref{Fig:Mech}-(a) reads,
\begin{eqnarray}
\mathcal{M}_{\mathrm{Dir}} &=& \Big[\bar{v}(p_2,m_e) (ie \gamma^\mu ) u(p_1,m_e)\Big] \frac{-g_{\mu \nu}}{q^2} \Big[ie (p_4^\nu -p_3^\nu) \mathcal{F}_K(q^2)\Big].\nonumber\\
\end{eqnarray}
with $q=k_1 +k_2$.It should be noted that in Eq.~\eqref{Eq:Lag1}, the kaon is considered as a point-like particle. In the present estimation, we include a form factor in the amplitude to depict the internal structure of kaon. In principle, the form factors of the hadrons in the time-like region are complex~\cite{Yang:2019mzq, Chen:2008hka}.  Fortunately, the argument of the form factor changes slowly when $q^2$ is far away from the threshold. Thus, in the present work, we suppose that the argument of the kaon form factor is approximately to be a constant in the considered center-of-mass energy range. Consequently, the argument of the form factor can be absorbed by the phase angle between two amplitudes and the form factor can be treated as a real function of $s$ and the concrete form of the form factor adopted in the present work is supposed to be, 
\begin{eqnarray}
\mathcal{F}_K(s) = a s^b e^{-cs} 	
\end{eqnarray}
with $s=q^2$, and $a$, $b$, $c$ are considered as free parameters, which could be determined by fitting the experimental data.  

The relevant effective Lagrangians involved in  Fig.~\ref{Fig:Mech}-(b) read~\cite{Chen:2011cj,Bauer:1975bw, Bauer:1975bv, Bauer:1977iq},
\begin{eqnarray}
\mathcal{L}_{VKK} &=& i g_{VKK} V^\mu (\bar{K} \partial_\mu K-\partial_\mu \bar{K} K), \nonumber\\
\mathcal{L}_{\gamma V} &=&-e \frac{m_V^2}{f_V} V^\mu A_\mu, \label{Eq:Lag}
\end{eqnarray}
and the corresponding amplitudes is, 
\begin{eqnarray}
\mathcal{M}_{\mathrm{Res}} &=& \Big[\bar{v}(p_2,m_e) (ie \gamma_\mu ) u(p_2,m_e)\Big] \frac{-g^{\mu \rho}}{q^2}\Big(-e \frac{m_V^2}{f_V} \Big) \nonumber\\
&&\frac{-g_{\rho \nu}+q_\rho q_\nu/m_V^2}{q^2-m_V^2+i m_V \Gamma_V}  \Big[ig_{VKK} (p_4^\nu -p_3^\nu) \Big].
\end{eqnarray} 

Then, the total amplitude of $e^+e^- \to K \bar{K}$ is,
\begin{eqnarray}
\mathcal{M}_{\mathrm{Tot}}&=&\mathcal{M}_{\mathrm{Dir}}+e^{i\phi} \mathcal{M}_{\mathrm{Res}}\nonumber\\
&=&\Big[\bar{v}(p_2,m_e) (ie \gamma^\mu ) u(p_2,m_e)\Big] \frac{1}{q^2} \Big[ie (p_4^\nu -p_3^\nu) \Big]\nonumber\\
&&\times \Big[-F_K(q^2)g_{\mu \nu} -e^{i\phi} g_V \frac{-g_{\mu \nu }+q_\mu q_\rho/m_V^2}{q^2-m_V^2+im_V\Gamma_V}  \Big]
\end{eqnarray}
where $\phi$ is the phase angle between two amplitudes and $g_V= g_{V KK} m_V^2/f_V$.

\begin{figure*}[htb]
\centering
\scalebox{1.1}{\includegraphics{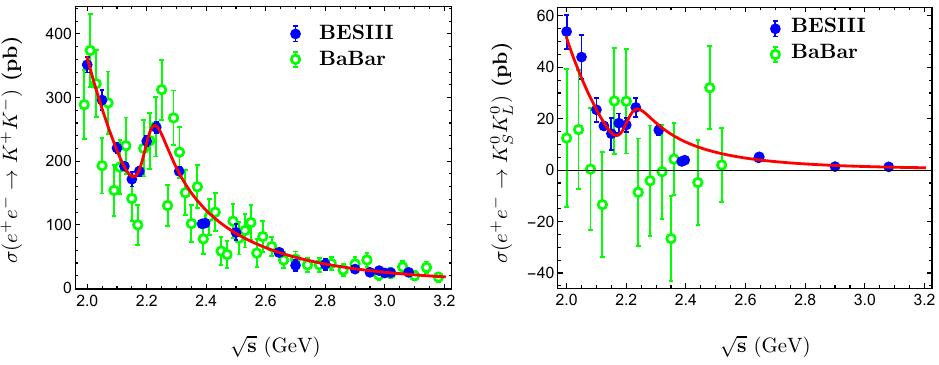}}\caption{(Color online.) Combined fit to the cross sections for $e^+ e^- \to K^+ K^-$ (left panel) and $e^+e^- \to K_S^0 K_L^0$ (right panel) with free isospin factor $R$. For comparison, the experimental data from the BESIII Collaboration~\cite{BESIII:2018ldc,BESIII:2021yam} and BaBar Collaboration~\cite{BaBar:2013jqz} are also presented. \label{Fig:freeR}}
\end{figure*}
\begin{figure*}[htb]
\centering
\scalebox{1.05}{\includegraphics{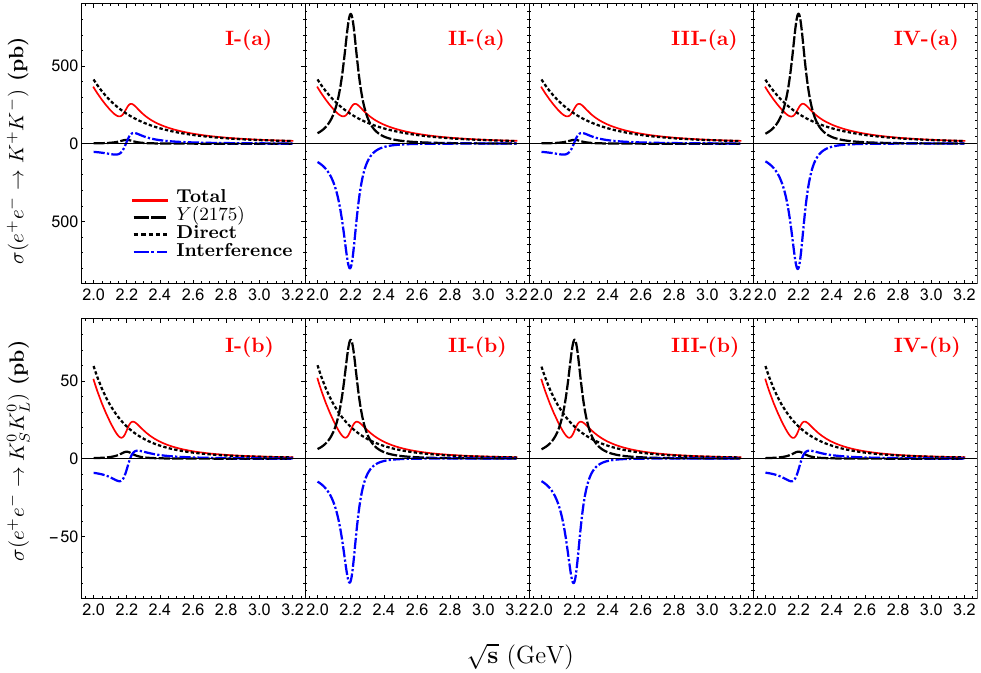}}	
\caption{(Color online.) Different solutions for individual contributions to the cross sections for $e^+ e^- \to K^+ K^-$ (up panels) and $e^+e^- \to K_S^0 K_L^0$ (low panels) with free isospin factor $R$. The red solid curves are our fits to the cross sections for $e^+ e^- \to K^+ K^-$ and $e^+e^- \to K_S^0 K_L^0$, while the black dashed, black dotted, and blue dash-dotted curves are the contributions from direct coupling process, resonance intermediate process, and their interferences, respectively. \label{Fig:freeR-Ind} }
\end{figure*}

\begin{table*}[htb]
\caption{The values of the parameters obtained by fitting the cross sections for $e^+ e^- \to K^+ K^-$ and $e^+ e^- \to K_S^0 K_L^0$ from the BESIII Collaboration~\cite{BESIII:2018ldc,BESIII:2021yam}\label{Tab:Para-freeR}}
\begin{tabular}{p{3cm}<\centering p{1.5cm}<\centering p{3cm}<\centering p{3cm}<\centering p{3cm}<\centering p{3cm}<\centering}
\toprule[1pt]
& parameter & Solutions I & Solutions II &Solutions III &Solutions IV \\
\midrule[1pt]
\multirow{2}{*}{Resonance parameters}
                     & $m_V$ (MeV)  & \multicolumn{4}{c}{$2200.9\pm 5.5$}\\
                     & $\Gamma_V$ (MeV) &\multicolumn{4}{c}{$107.1\pm 11.0$} \\
\midrule[1pt]
\multirow{5}{*}{Charged channel}      
 &$a$ & $  5.28 \pm 0.19 $ & $  5.52 \pm 0.65 $ & $  5.30 \pm 0.20 $ & $  5.20 \pm  3.74 $\\
&$b$ & $ -4.92 \pm 0.01 $ & $ -5.05 \pm 0.32 $ & $ -4.94 \pm 0.01 $ & $ -4.90 \pm  0.49 $\\  
&$c$ & $ -0.17 \pm 0.01 $ & $ -0.18 \pm 0.02 $ & $ -0.17 \pm 0.01 $ & $ -0.17 \pm  0.04 $\\
&$g_V $& $  (2.03 \pm 0.23)\times 10^{-2} $ & $  (1.20 \pm 0.12)\times 10^{-1} $ & $  (2.03\pm 0.23)\times 10^{-2}  $ & $  (1.17 \pm 0.01) \times 10^{-1} $\\
&$\phi$ & $  3.04 \pm 0.11 $ & $  4.62 \pm 0.01 $ & $  3.04 \pm 0.11 $ & $  4.62 \pm 0.01  $\\  
\midrule[1pt]                    
\multirow{5}{*}{Neutral channel}     
 &$a^\prime$ & $  6.86 \pm 0.87 $ & $  7.86 \pm 2.31 $ & $  6.86 \pm 0.85 $ & $  6.92 \pm 0.84 $\\
&$b^\prime$ & $ -7.16 \pm 0.02 $ & $ -7.52 \pm 0.70 $ & $ -7.17 \pm 0.03 $ & $ -7.19 \pm 0.09 $ \\ 
&$c^\prime$ & $ -0.25 \pm 0.02 $ & $ -0.28 \pm 0.65 $ & $ -0.25 \pm 0.02 $ & $ -0.25 \pm 0.02 $  \\
&$R$ & $  0.43 \pm 0.08 $ & $  0.30 \pm 0.01 $ & $  1.77 \pm 0.11 $ & $  0.07 \pm 0.01 $\\
&$\phi^\prime$& $  3.50 \pm 0.19 $ & $  4.59 \pm 0.04 $ & $  4.59 \pm 0.04 $ & $  3.50 \pm 0.16 $\\                           
\bottomrule[1pt]
\end{tabular}
\end{table*}

With the above amplitude, we can get the cross sections for $e^+e^- \to K \bar{K}$. In the present scheme, there are three kinds of free parameters, which are,
\begin{itemize}
	\item  The resonances parameters of the involved vector mesons $m_V$, $\Gamma_V$, which are the same for both $e^+ e^- \to K^+ K^-$ and $e^+ e^- \to K_S^0 K_L^0$ processes. 
	\item $a$, $b$, $c$, $\phi$, $g_V=g_{VK^+ K^-} m_V^2/f_V$, which are involved only in the $e^+ e^- \to K^+ K^-$ process.
	\item $a^\prime$, $b^\prime$, $c^\prime$, $\phi^\prime$, $R=g_V^\prime/g_V$ with $g_V^\prime=g_{VK^0 \bar{K}^0} m_V^2/f_V$, which are involved only in the $e^+e^-\to K_S^0 K_L^0$ process. 
\end{itemize}

\section{Numerical Results \label{Sec:Num}}
As indicated Ref~\cite{Chen:2020xho}, the uncertainties of the cross sections for $e^+ e^- \to K^+ K^-$ and $e^+ e^- \to K_S^0 K_L^0$ from the BESIII Collaboration are rather small, however, the continuity of the data is not as good as their precision~\cite{BESIII:2018ldc,BESIII:2021yam}. Particularly, there are only two nearly degenerated data point in the range $\sqrt{s}=2.31\sim 2.5$ GeV, which can not reflect more details of the cross sections in this energy range. In addition, some experimental measurement from BaBar, Belle and BESIII Collaborations~\cite{BaBar:2007ptr, BES:2007sqy, Belle:2008kuo, BESIII:2014ybv, Shen:2009mr}, and the theoretical investigations form Refs.~\cite{Chen:2018kuu} have indicated a vector resonance near 2.4 GeV, thus, we excluded these two data points in the present fit.

\subsection{Combined fit with free isospin factor}
Our fit to the cross sections for $e^+e^- \to K^+ K^-$ and $e^+ e^- \to K_S^0 K_L^0$ are presented in Fig.~\ref{Fig:freeR}, where the blue full circles with error bars and open circles with error bars are experimental data from BESIII~\cite{BESIII:2018ldc,BESIII:2021yam} and BaBar Collaborations~\cite{BaBar:2013jqz}, respectively, while the red solid lines are our fit the cross sections in the present work. From the figure one can find that the cross sections could be well reproduced except  for the data around 2.4 GeV, which have been exclude in our fit as we clarified above. 

The parameters determined by fitting the cross sections for $e^+ e^- \to K^+ K^-$ and $e^+e^- \to K_S^0 K_L^0$ are listed in Table \ref{Tab:Para-freeR}. From the table, one can find that the resonance parameters are determined to be,
\begin{eqnarray}
m_V &=&  (2200.9 \pm 5.5) \ \mathrm{MeV}, \nonumber\\
\Gamma_V &=& (107.1 \pm 11.0) \ \mathrm{MeV}, 	
\end{eqnarray}
respectively, which is smaller than the individual fit from BESIII Collaboration in mass~\cite{BESIII:2018ldc,BESIII:2021yam}. Particularly, the fitted mass is differ from that of $e^+e^- \to K^+ K^-$ by about $2.6 \sigma$, and differ from that of $e^+e^- \to K_S^0 K_L^0$ by $3.5 \sigma$.

It should be noted that in the interference scenario, the interference between the direct coupling process and the resonance intermediate process could be constructive or destructive for $e^+ e^- \to K^+ K^-$ or $e^+ e^- \to K_S^0 K_L^0$ process. Thus, one should have four different solutions for the parameters in the combined analysis. The individual contributions for these four solutions are presented in Fig.~\ref{Fig:freeR-Ind}, where the interferences in diagram I-(a), III-(a), I-(b) and IV-(b) are constructive in the cross sections peak position, while in II-(a), VI-(a),  II-(b), and III-(b), the interferences are destructive. Thus, in these four solutions, the isospin factor $R$ will be different.

Since the mass of $\phi(2170)$ is much greater than the $K\bar{K}$ threshold, thus, the isospin factor $R$ is expected to be one under isospin symmetry. However, as shown in Table~\ref{Tab:Para-freeR}, our combined analysis indicate that the isospin factor $R$ is fitted to be $0.43 \pm 0.08$, $0.30 \pm 0.01$, $1.77 \pm 0.11$, and $0.07 \pm 0.01$ for different solution, and none of them is consistent with the expectation of isospin symmetry. This phenomena indicate that there may exist other vector resonance near $\phi(2170)$, since both isospin vector state, such as $\rho$ meson, and isospin scalar state, such as $\omega$ and $\phi$ meson, can contribute to this process. In addition, the fitted isospin factor varies in a large range indicated another possibility that the experimental data are overfitted.

\begin{figure*}[t]
\centering
\scalebox{1.05}{\includegraphics{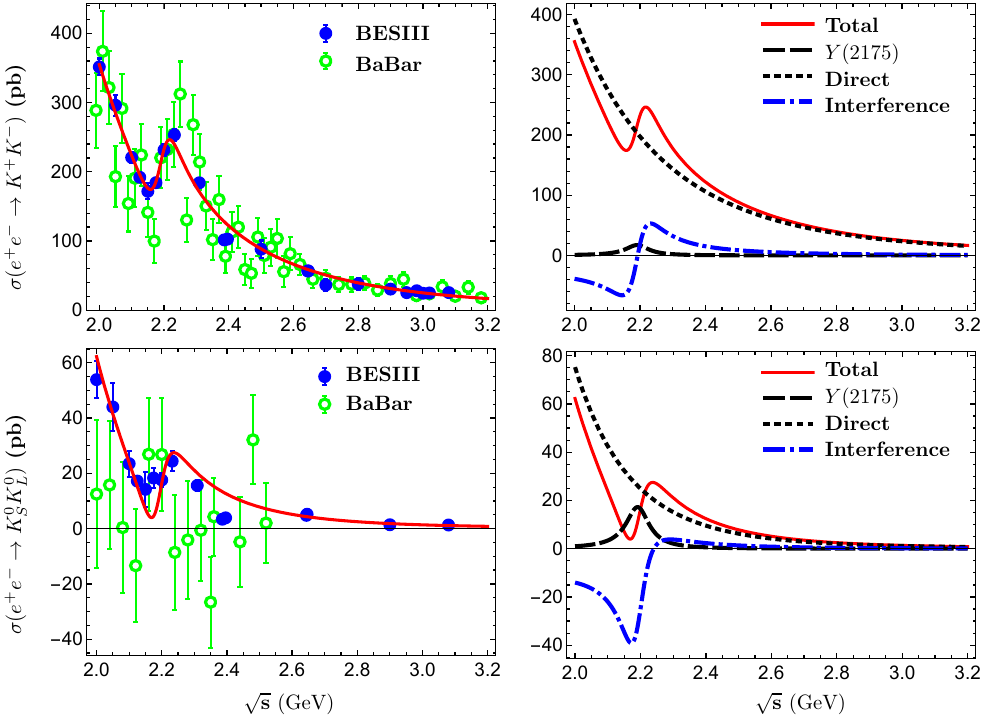}}
\caption{(Color online.) Combined fit to the cross sections for $e^+ e^- \to K^+ K^-$ and $e^+e^- \to K_S^0 K_L^0$ with fixed isospin factor $R$. The left panels are the fitted cross sections comparing the experimental data, while the right panels are the individual contributions from the direct coupling process (black dashed curves), the resonance intermediate process (black dashed curves) and their interferences (blue dash-dotted curves).\label{Fig:FixedR}}
\end{figure*}

\subsection{Combined fit with fixed isospin factor}
As indicate in the above discussions that the fitted values of the isospin factor varies from much smaller than 1 to 1.77. To further check the reason of the isospin symmetry breaking, we fix the isospin factor to be one and then fit the experimental data in the following. With the fixed isospin factor, we obtain the mass and width of the vector resonance to be,
\begin{eqnarray}
m&=& (2191.6 \pm 5.0 )\ \mathrm{MeV},\nonumber\\
\Gamma &=& (86.1 \pm 9.2)\ \mathrm{MeV},
\end{eqnarray}
respectively, which are differ from the ones obtained with free isospin factor by 1.3$\sigma$ in mass and $1.5\sigma$ in width. 

\begin{table}
\caption{The values of the parameters except for the resonance parameters obtained by fitting the cross sections for $e^+ e^- \to K^+ K^-$ and $e^+ e^- \to K_S^0 K_L^0$ from the BESIII Collaboration~\cite{BESIII:2018ldc,BESIII:2021yam} with fixed isospin factor. \label{Tab:Para-fixedR}}
\begin{tabular}{p{1.5cm}<\centering p{2.5cm}<\centering p{1.5cm}<\centering p{2.5cm}<\centering}
\toprule[1pt]
 \multicolumn{2}{c}{$e^+ e^- \to K^+ K^-$} &  \multicolumn{2}{c}{$e^+ e^- \to K_S^0 K_L^0$}\\
Paramter & Value & Paramter & Value\\
\midrule[1pt]

$a$    & $  3.20\pm 0.11$                  &$a^\prime$    & $  8.60\pm 0.85$\\
$b$    & $ -3.70\pm 0.01$                  &$b^\prime$    & $ -7.14\pm 0.04$\\
$c$    & $ -0.08\pm 0.01$                  &$c^\prime$    & $ -0.22\pm 0.02$\\
$g_V$  & $  (1.37\pm 0.14) \times 10^{-2}$ & $R$          & $1.00$ (fixed)\\
$\phi$ & $  3.17 \pm 0.12    $             & $\phi^\prime$ & $  4.00\pm 0.12$\\
\bottomrule[1pt]
\end{tabular}	
\end{table}

The fitted cross sections and the individual contributions to the processes $e^+ e^-\to K^+ K^-$ and $e^+ e^- \to K_S^0 K_L^0$ are presented in Fig.~\ref{Fig:FixedR}. From the figure one can find that with the parameters in Table.~\ref{Tab:Para-fixedR}, we can accurately fit the observed cross sections for $e^+ e^-\to K^+ K^-$, while the same model parameters fail to adequately reproduce the observed cross sections for the process $e^+ e^- \to K_S^0 K_L^0$, which indicate that the experimental data is not overfitted and there should be contributions from other vector meson states. In the vicinity of 2.2 GeV, there are several observed vector resonance, including $\rho(2150)$, $\omega(2220)$, which should also contribute to the process $e^+e^- \to K^+K^-$ and $e^+ e^- \to K_S^0 K_L^0$. More precise experimental measurement below 2.0 GeV and between $2.3\sim 2.5$ GeV may provide more information about the vector resonances in this energy range, which should be accessible for the BESIII Collaboration.

\section{Summary}
The observations of the strangeonium-like state $\phi(2170)$ have stimulated theorists' great interests in its internal structure since the similar observed channels as $Y(4260)$ and $Y_b(10890)$. The exotic hadron interpretations, such as $ss\bar{s}\bar{s}$ or $su\bar{s} \bar{u}$ tetraquark, $\Lambda\bar{\Lambda}$ or $\phi f_0(980)$ molecular and strangeonium hybrid, have been proposed. In addition to the exotic hadron interpretations, $\phi(2170)$ has also been interpreted in the conventional strangeonium frame, where $\phi(2170)$ was considered as $3^3S_1$ or  $2^3D_1$ state. The theoretical investigations indicate that the open strange decay modes were crucial to distinguish different interpretations. 

Recently, the BESIII Collaboration have reported their measurements of the cross sections for $e^+e^- \to K^+ K^-$ and $e^+ e^- \to K_S^0 K_L^0$, and clear structure near 2.2 GeV were reported. However, the measured mass and width in these two channels are much different with the resonance parameters of $\phi(2170)$. In addition, the experimental measurements indicate large isospin symmetry breaking in the resonance region. In the present work, we perform a combined analysis of the cross sections for $e^+ e^- \to K^+ K^-$ and $e^+ e^- \to K_S^0 K_L^0$, where the interferences between the direct coupling process and the resonance intermediate process are taken into consideration. When setting the isospin factor as a free parameter, we can well reproduce the structures in the cross sections for $e^+ e^- \to K^+ K^-$ and $e^+ e^- \to K_S^0 K_L^0$, however, the    isospin factor deviates with the isospin symmetry conservation predicted value of one. 

To further check the reason of the isospin symmetry breaking, we fix the isospin factor to be one and then fit the experimental data. Our estimations indicate that we can accurately fit the observed cross sections for $e^+ e^-\to K^+ K^-$, while the same model parameters fail to adequately reproduce the observed cross section for the process $e^+ e^- \to K_S^0 K_L^0$, which indicate that there should be contributions from other vector meson states, such as $\rho(2150)$ and $\omega(2220)$. More precise experimental measurement below 2.0 GeV and between $2.3\sim 2.5$ GeV may provide more information about the vector resonances in this energy range, which should be accessible for the BESIII Collaboration.

\section{ACKNOWLEDGMENTS} This work is partly supported by the National Natural Science Foundation of China under the Grant Nos. 12175037, 12335001 and 12135005, as well as supported, in part, by National Key Research and Development Program under the contract No. 2024YFA1610503

%\clearpage\newpage
\bibliographystyle{unsrt}
\bibliography{Y2175.bib}

\begin{thebibliography}{10}

\bibitem{BaBar:2006gsq}
Bernard Aubert et~al.
\newblock {A Structure at 2175-MeV in $e^{+} e^{-} \to \phi$ f0(980) Observed
  via Initial-State Radiation}.
\newblock {\em Phys. Rev. D}, 74:091103, 2006.

\bibitem{BES:2007sqy}
Medina Ablikim et~al.
\newblock {Observation of $Y(2175)$ in $J / \psi \to \eta \phi f_0(980)$}.
\newblock {\em Phys. Rev. Lett.}, 100:102003, 2008.

\bibitem{Belle:2008kuo}
C.~P. Shen et~al.
\newblock {Observation of the $\phi(1680)$ and the $Y(2175)$ in $e^+e^-\to \phi
  \pi^+ \pi^-$}.
\newblock {\em Phys. Rev. D}, 80:031101, 2009.

\bibitem{BaBar:2011btv}
J.~P. Lees et~al.
\newblock {Cross Sections for the Reactions $e^+e^- \to K^+ K^- \pi^+ \pi^-$,
  $K^+ K^- \pi^0 \pi^0$, and $K^+ K^- K^+ K^-$ Measured Using Initial-State
  Radiation Events}.
\newblock {\em Phys. Rev. D}, 86:012008, 2012.

\bibitem{BaBar:2007ceh}
Bernard Aubert et~al.
\newblock {Measurements of $e^{+} e^{-} \to K^{+} K^{-} \eta$, $K^{+} K^{-}
  \pi^0$ and $K^0_{s} K^\pm \pi^\mp$ cross- sections using initial state
  radiation events}.
\newblock {\em Phys. Rev. D}, 77:092002, 2008.

\bibitem{BaBar:2007ptr}
Bernard Aubert et~al.
\newblock {The $e^+ e^- \to K^+ K^- \pi^+ \pi^-$, $K^+ K^- \pi^0 \pi^0$ and
  $K^+ K^- K^+ K^-$ cross-sections measured with initial-state radiation}.
\newblock {\em Phys. Rev. D}, 76:012008, 2007.

\bibitem{BESIII:2017qkh}
Medina Ablikim et~al.
\newblock {Observation of $e^+ e^- \to \eta Y(2175)$ at center-of-mass energies
  above 3.7 GeV}.
\newblock {\em Phys. Rev. D}, 99(1):012014, 2019.

\bibitem{BESIII:2020vtu}
M.~Ablikim et~al.
\newblock {Observation of a Resonant Structure in $e^{+}e^{-} \to
  K^{+}K^{-}\pi^{0}\pi^{0}$}.
\newblock {\em Phys. Rev. Lett.}, 124(11):112001, 2020.

\bibitem{BESIII:2020gnc}
M.~Ablikim et~al.
\newblock {Observation of a structure in $e^{+}e^{-} \to \phi \eta^{\prime}$ at
  $\sqrt{s}$ from 2.05 to 3.08 GeV}.
\newblock {\em Phys. Rev. D}, 102(1):012008, 2020.

\bibitem{BESIII:2020xmw}
M.~Ablikim et~al.
\newblock {Observation of a resonant structure in $e^{+}e^{-} \to \omega\eta$
  and another in $e^{+}e^{-} \to \omega\pi^{0}$ at center-of-mass energies
  between 2.00 and 3.08 GeV}.
\newblock {\em Phys. Lett. B}, 813:136059, 2021.

\bibitem{BESIII:2021bjn}
Medina Ablikim et~al.
\newblock {Study of the process $e^{+}e^{-}\rightarrow\phi\eta$ at
  center-of-mass energies between 2.00 and 3.08 GeV}.
\newblock {\em Phys. Rev. D}, 104(3):032007, 2021.

\bibitem{BESIII:2022wxz}
M.~Ablikim et~al.
\newblock {Measurement of $e^{+}e^{-} \to K^{+}K^{-}\pi^{0}$ cross section and
  observation of a resonant structure}.
\newblock {\em JHEP}, 07:045, 2022.

\bibitem{BESIII:2018ldc}
M.~Ablikim et~al.
\newblock {Measurement of $e^{+} e^{-} \rightarrow K^{+} K^{-}$ cross section
  at $\sqrt{s} = 2.00 - 3.08$ GeV}.
\newblock {\em Phys. Rev. D}, 99(3):032001, 2019.

\bibitem{BESIII:2021yam}
Medina Ablikim et~al.
\newblock {Cross section measurement of $e^{+}e^{-} \to K_{S}^{0}K_{L}^{0}$ at
  $\sqrt{s}=2.00-3.08~{GeV}$}.
\newblock {\em Phys. Rev. D}, 104(9):092014, 2021.

\bibitem{BESIII:2021aet}
Medina Ablikim et~al.
\newblock {Measurement of
  e+e-\textrightarrow{}\ensuremath{\phi}\ensuremath{\pi}+\ensuremath{\pi}-
  cross sections at center-of-mass energies from 2.00 to 3.08~GeV}.
\newblock {\em Phys. Rev. D}, 108(3):032011, 2023.

\bibitem{BESIII:2023xac}
Medina Ablikim et~al.
\newblock {Measurement of the e$^{+}$e$^{-}$ \textrightarrow{} $ {K}_S^0{K}_L^0
  $\ensuremath{\pi}$^{0}$ cross sections from $ \sqrt{s} $ = 2.000 to 3.080
  GeV}.
\newblock {\em JHEP}, 01:180, 2024.

\bibitem{ParticleDataGroup:2024cfk}
S.~Navas et~al.
\newblock {Review of particle physics}.
\newblock {\em Phys. Rev. D}, 110(3):030001, 2024.

\bibitem{ParticleDataGroup:2018ovx}
M.~Tanabashi et~al.
\newblock {Review of Particle Physics}.
\newblock {\em Phys. Rev. D}, 98(3):030001, 2018.

\bibitem{BESIII:2014ybv}
M.~Ablikim et~al.
\newblock {Study of $J/\psi \to \eta \phi \pi^+ \pi^-$ at BESIII}.
\newblock {\em Phys. Rev. D}, 91(5):052017, 2015.

\bibitem{BaBar:2005hhc}
Bernard Aubert et~al.
\newblock {Observation of a broad structure in the $\pi^+ \pi^- J/\psi$ mass
  spectrum around 4.26-GeV/c$^2$}.
\newblock {\em Phys. Rev. Lett.}, 95:142001, 2005.

\bibitem{BaBar:2006ait}
Bernard Aubert et~al.
\newblock {Evidence of a broad structure at an invariant mass of 4.32-
  $GeV/c^{2}$ in the reaction $e^{+} e^{-} \to \pi^{+} \pi^{-} \psi_{2S}$
  measured at BaBar}.
\newblock {\em Phys. Rev. Lett.}, 98:212001, 2007.

\bibitem{Belle:2007xek}
K.~F. Chen et~al.
\newblock {Observation of anomalous Upsilon(1S) pi+ pi- and Upsilon(2S) pi+ pi-
  production near the Upsilon(5S) resonance}.
\newblock {\em Phys. Rev. Lett.}, 100:112001, 2008.

\bibitem{Wang:2006ri}
Zhi-Gang Wang.
\newblock {Analysis of the Y(2175) as a tetraquark state with QCD sum rules}.
\newblock {\em Nucl. Phys. A}, 791:106--116, 2007.

\bibitem{Jiang:2023atq}
Yi-Wei Jiang, Wei-Han Tan, Hua-Xing Chen, and Er-Liang Cui.
\newblock {Strong Decays of the $\phi$(2170) as a Fully Strange Tetraquark
  State}.
\newblock {\em Symmetry}, 16(8):1021, 2024.

\bibitem{Agaev:2019coa}
S.~S. Agaev, K.~Azizi, and H.~Sundu.
\newblock {Nature of the vector resonance $Y(2175)$}.
\newblock {\em Phys. Rev. D}, 101(7):074012, 2020.

\bibitem{Drenska:2008gr}
N.~V. Drenska, R.~Faccini, and A.~D. Polosa.
\newblock {Higher Tetraquark Particles}.
\newblock {\em Phys. Lett. B}, 669:160--166, 2008.

\bibitem{Deng:2010zzd}
Chengrong Deng, Jialun Ping, Fan Wang, and T.~Goldman.
\newblock {Tetraquark state and multibody interaction}.
\newblock {\em Phys. Rev. D}, 82:074001, 2010.

\bibitem{Ding:2006ya}
Gui-Jun Ding and Mu-Lin Yan.
\newblock {A Candidate for 1-- strangeonium hybrid}.
\newblock {\em Phys. Lett. B}, 650:390--400, 2007.

\bibitem{Ding:2007pc}
Gui-Jun Ding and Mu-Lin Yan.
\newblock {Y(2175): Distinguish Hybrid State from Higher Quarkonium}.
\newblock {\em Phys. Lett. B}, 657:49--54, 2007.

\bibitem{Dudek:2011bn}
Jozef~J. Dudek.
\newblock {The lightest hybrid meson supermultiplet in QCD}.
\newblock {\em Phys. Rev. D}, 84:074023, 2011.

\bibitem{Ho:2019org}
J.~Ho, R.~Berg, T.~G. Steele, W.~Chen, and D.~Harnett.
\newblock {Is the $Y(2175)$ a Strangeonium Hybrid Meson?}
\newblock {\em Phys. Rev. D}, 100(3):034012, 2019.

\bibitem{Klempt:2007cp}
Eberhard Klempt and Alexander Zaitsev.
\newblock {Glueballs, Hybrids, Multiquarks. Experimental facts versus QCD
  inspired concepts}.
\newblock {\em Phys. Rept.}, 454:1--202, 2007.

\bibitem{Dong:2017rmg}
Yubing Dong, Amand Faessler, Thomas Gutsche, Qifang L\"u, and Valery~E.
  Lyubovitskij.
\newblock {Selected strong decays of $\eta(2225)$ and $\phi(2170)$ as $\Lambda
  \bar\Lambda$ bound states}.
\newblock {\em Phys. Rev. D}, 96(7):074027, 2017.

\bibitem{MartinezTorres:2008gy}
A.~Martinez~Torres, K.~P. Khemchandani, L.~S. Geng, M.~Napsuciale, and E.~Oset.
\newblock {The X(2175) as a resonant state of the phi K anti-K system}.
\newblock {\em Phys. Rev. D}, 78:074031, 2008.

\bibitem{Barnes:2002mu}
T.~Barnes, N.~Black, and P.~R. Page.
\newblock {Strong decays of strange quarkonia}.
\newblock {\em Phys. Rev. D}, 68:054014, 2003.

\bibitem{Wang:2012wa}
Xiao Wang, Zhi-Feng Sun, Dian-Yong Chen, Xiang Liu, and Takayuki Matsuki.
\newblock {Non-strange partner of strangeonium-like state Y(2175)}.
\newblock {\em Phys. Rev. D}, 85:074024, 2012.

\bibitem{Jin:2019gfa}
Yi~Jin, Shi-Yuan Li, Yan-Rui Liu, Zhao-Xia Meng, Zong-Guo Si, and Tao Yao.
\newblock {Exclusive production ratio of neutral to charged kaon pair in $e^+
  e^-$ annihilation continuum via a relativistic quark model}.
\newblock {\em Phys. Rev. C}, 102(1):015201, 2020.

\bibitem{Chen:2020xho}
Dian-Yong Chen, Jing Liu, and Jun He.
\newblock {Reconciling the $X$(2240) with the $Y$(2175)}.
\newblock {\em Phys. Rev. D}, 101(7):074045, 2020.

\bibitem{Yang:2019mzq}
Yongliang Yang, Dian-Yong Chen, and Zhun Lu.
\newblock {Electromagnetic form factors of $\Lambda$ hyperon in the vector
  meson dominance model}.
\newblock {\em Phys. Rev. D}, 100(7):073007, 2019.

\bibitem{Chen:2008hka}
D.~Y. Chen, H.~Q. Zhou, and Y.~B. Dong.
\newblock {Two-Photon Exchange Contribution to Proton Form Factors in Time-Like
  region}.
\newblock {\em Phys. Rev. C}, 78:045208, 2008.

\bibitem{Chen:2011cj}
Dian-Yong Chen, Xiang Liu, and Takayuki Matsuki.
\newblock {Two Charged Strangeonium-Like Structures Observable in the $Y(2175)
  \to \phi(1020)\pi^{+} \pi^{-}$ Process}.
\newblock {\em Eur. Phys. J. C}, 72:2008, 2012.

\bibitem{Bauer:1975bw}
T.~Bauer and D.~R. Yennie.
\newblock {Corrections to Diagonal VDM in the Photoproduction of Vector Mesons.
  2. Phi-omega Mixing}.
\newblock {\em Phys. Lett. B}, 60:169--171, 1976.

\bibitem{Bauer:1975bv}
T.~Bauer and D.~R. Yennie.
\newblock {Corrections to VDM in the Photoproduction of Vector Mesons. 1. Mass
  Dependence of Amplitudes}.
\newblock {\em Phys. Lett. B}, 60:165--168, 1976.

\bibitem{Bauer:1977iq}
T.~H. Bauer, R.~D. Spital, D.~R. Yennie, and F.~M. Pipkin.
\newblock {The Hadronic Properties of the Photon in High-Energy Interactions}.
\newblock {\em Rev. Mod. Phys.}, 50:261, 1978.
\newblock [Erratum: Rev.Mod.Phys. 51, 407 (1979)].

\bibitem{BaBar:2013jqz}
J.~P. Lees et~al.
\newblock {Precision measurement of the $e^+e^- \to K^+K^-(\gamma)$ cross
  section with the initial-state radiation method at BABAR}.
\newblock {\em Phys. Rev. D}, 88(3):032013, 2013.

\bibitem{Shen:2009mr}
C.~P. Shen and C.~Z. Yuan.
\newblock {Combined fit to BaBar and Belle Data on $e^+ e^- \to \phi \pi^+
  \pi^-$ and $\phi f_0(980)$}.
\newblock {\em Chin. Phys. C}, 34:1045--1051, 2010.

\bibitem{Chen:2018kuu}
Hua-Xing Chen, Cheng-Ping Shen, and Shi-Lin Zhu.
\newblock {A possible partner state of the $Y(2175)$}.
\newblock {\em Phys. Rev. D}, 98(1):014011, 2018.

\end{thebibliography}
\end{document}